\documentclass[12pt]{article}

\begin{document}

\hyphenation{Min-kow-ski} \hyphenation{cosmo-logical} %
\hyphenation{holo-graphy} \hyphenation{super-symmetry} %
\hyphenation{super-symmetric}

\rightline{VPI-IPNAS-08-07} 
\centerline{\Large \bf }
\vskip0.25cm 
\centerline{\Large \bf }
\vskip0.25cm %
\centerline{\Large \bf High Temperature Superconductivity and Effective Gravity}
\vskip0.25cm 
\vskip 1cm

\renewcommand{\thefootnote}{\fnsymbol{footnote}} 
\centerline{{\bf Djordje Minic\footnote{dminic@vt.edu} and Jean J.
Heremans\footnote{heremans@vt.edu}}} 
\vskip .5cm 
\centerline{\it Department of Physics, Virginia Tech} 
\centerline{\it Blacksburg, VA 24061, U.S.A.}

\begin{abstract}
We argue that an approach involving effective gravity could play a crucial
role in elucidating the properties of the high temperature superconducting
materials. In particular we propose that the high critical temperature might
be naturally explained in a framework constructed as a direct condensed
matter analog of the Randall-Sundrum approach to a geometrization of the
hierarchy problem in high-energy physics.
\end{abstract}

\vskip .5cm

\setcounter{footnote}{0} \renewcommand{\thefootnote}{\arabic{footnote}}

\newpage

\section{Introduction}

The problem of high temperature superconductivity is one of the most
outstanding puzzles in contemporary physics \cite{htcreview}. Recently much
activity has been devoted to the possible relevance of the gauge
theory/gravity duality (i.e. the anti de Sitter space/conformal field theory
correspondence) to this remarkable problem \cite{adscft}. The general
concept of duality, on which a lot of recent progress in quantum field
theory and string theory hinges, has also been used recently in various
condensed matter settings \cite{sdfqhe, son}. Furthermore, there has been a lot
of work regarding the induced gravitational processes in various condensed
matter systems (see for example \cite{abh}, \cite{volovik} for some recent
samplings of the literature).

In this note we observe that if indeed effective gravity could arise in
condensed matter systems then there are possible dramatic effects associated
with its geometric physics. For example, widely separated energy scales can
be related through a warping (red-shift) factor. This possibility has been
widely discussed in the recent high-energy literature. In particular, in the
context of the Randall-Sundrum scenario \cite{rs}, a geometric warping
factor is used to naturally bump up the 1TeV scale to the Planck scale, thus
evading the hierarchy problem. 

The main aim of this note is to point out a close analogy between the
naturalness problem associated with the Higgs mass from the realm of
particle physics and the naturalness problem associated with the high
critical temperature ($T_{C}$) of high temperature superconductors. In view
of the recent geometric proposals for dealing with the particle physics
naturalness problem, i.e. the Randall-Sundrum scenarios (which indeed have
not been tested experimentally, and are, at the moment, only theoretical
suggestions) we point out, that a similar geometric mechanism can be
naturally used in a condensed matter context. We propose a condensed matter
analog of the Randall-Sundrum scenario for a geometrically induced hierarchy
of scales. We argue that in the context of superconductivity, such a
scenario can lead to the high $T_{C}$ of cuprate (Cu-O) and oxypnictide
(Fe-As) superconductors, both representing families of layered anisotropic
compounds with complex normal-state Fermi surfaces. Our discussion is based
entirely on effective field theory (i.e. a gravitationally dressed
Landau-Ginsburg approach) and does not address deeper issues concerning the
microscopic mechanism of high $T_{C}$ superconductivity. Nevertheless, the
new effective description may point toward previously unexplored avenues for
understanding the underlying microscopic physics. Because our approach
concentrates on the subsection of the phase diagram containing the strange
metal/superconducting phase, we ask how the usual canonical Wilsonian
thinking associated with naturalness of Fermi liquids should be modified for
a geometric scenario akin to Randall-Sundrum to be realized in an effective
Landau-Ginsburg description of the superconducting phase. Thus our proposal
can be summarized as: a) a description of high $T_{C}$ superconductors based
on a gravitationally dressed version of the canonical Landau-Ginsburg
Lagrangian (and the gravitationally dressed Fermi liquid as its normal
phase) presented in Section 2 and b) a specific proposal for the geometry of
this induced gravitational background, which is appropriate for the
anisotropic geometric properties of the known high $T_{C}$ materials
presented in Sections 3 and 4. The main result of the paper is presented in
Section 4. The specific geometry, as in the Randall-Sundrum scenario, uses a
slice of anti de Sitter space (AdS) and that is the gravitational metric
that appears in the gravitationally dressed Landau-Ginsburg Lagrangian for
the high-temperature superconductors.

Note that, even though a comparison can be drawn with the approach used for
the currently pursued AdS superconductors \cite{adscft} (which indeed build
on an AdS black hole, more specifically the Abelian Higgs model
in that background) we do not rely on the same gravitational description. Our
background, like the original Randall-Sundrum description, represents a
slice of the AdS space. A dual description of our proposal can be in
principle envisioned, and it is natural to conclude that this would involve
a non-relativistic AdS/conformal field theory correspondence \cite{son}.

\section{Fermi liquids vs. non-Fermi liquids}

As noticed by many authors, various condensed matter systems, such as
superfluids or Bose-Einstein condensates, exhibit physical phenomena that
can be interpreted by invoking effective gravity \cite{abh}. In these
contexts, which can be viewed as analog models of gravity, one speaks for
example of an effective acoustic metric, acoustic black holes (i.e. ``dumb
holes'') or emergent relativity.

Given the intricacies of the physics of high temperature superconducting
materials and the complexity of their phase diagram, one might ask whether
the strong electron correlations responsible for these phenomena can induce
effective gravitational effects, thus opening a possibility for a geometric
explanation of some of the outstanding puzzles.

We concentrate on one such puzzle: the actual high value of the critical
temperature. In particular, given the fact that in recent years high energy
physicists have invoked gravity to explain the unnatural mass scale
hierarchy we propose a condensed matter analog of the same mechanism. In the
context of high temperature superconductivity we argue for a geometric
reinterpretation of the hierarchy that exists between the ordinary critical
temperatures for low-temperature superconductors and their high temperature
relatives. This geometric reinterpretation may point to a new effective
gravitational physics which may be responsible, among other things, for the
high critical temperature of the high temperature superconducting materials.

A geometrically induced hierarchy of scales forces a revisiting of the
Wilsonian approach to the Fermi surface, to ascertain consequences for Fermi
Liquid theory in the renormalization group approach, and for the emergence
of non-Fermi Liquid behavior. From the effective field theory standpoint,
the question can be formulated: how does the low energy Wilsonian action
with effective gravity modify the usual scalings of the effective low energy
field theory of the Fermi surface?

\subsection{ The Fermi liquid}

To set the stage for our discussion let us summarize the classic effective
field theory of Landau Fermi liquids as most succinctly presented by
Polchinski \cite{joep}.

One starts with a \textit{natural} (i.e. not-finely tuned) Fermi surface and
decomposes the momenta into the Fermi momentum and a component orthogonal to
the Fermi surface 
\begin{equation}
\vec{p} = \vec{k} + \vec{l}
\end{equation}
and then one considers scaling of energy and momentum towards the Fermi
surface, in other words 
\begin{equation}
E \to sE, \quad \vec{k} \to \vec{k}, \quad \vec{l} \to s \vec{l}.
\end{equation}
The lowest order action, to quadratic order, is then given as 
\begin{equation}
S_{FL} = \int L_{FL} (\psi, \partial \psi) \equiv \int dt d^3 \vec{p} [ i
\psi^{*} (\vec{p}) \partial_t \psi (\vec{p})- (E(\vec{p}) -E_F(\vec{p}%
))\psi^{*} (\vec{p}) \psi (\vec{p})].
\end{equation}
Close to the Fermi surface 
\begin{equation}
E(\vec{p}) -E_F(\vec{p}) \sim l v_f, \quad v_F= \partial_{\vec{p}} E
\end{equation}
so that after the agreed renormalization orthogonal to the Fermi surface
(note that also $t \to s^{-1} t$) one obtains 
\begin{equation}
\psi \to s^{-1/2} \psi.
\end{equation}
Note that this scaling leads immediately to the usual two point function for
a free quasiparticle with a single particle pole. Now, by considering
4-Fermi interactions one can see that for generic momenta the 4-Fermi
interaction scales as a \textit{positive} power of $s$ and is thus
irrelevant at low energy. The measure over time contributes one negative
power, the measure over the momenta orthogonal to the Fermi surface
contributes 4 powers and the 4-Fermi interaction contributes $4/2$ negative
powers. The delta function over the 4 momenta generically does not scale. So
the overall generic scaling for the 4-Fermi vertex is indeed 
\begin{equation}
s^{-1+4-4/2} = s^{1}.
\end{equation}
This is valid, except if the momenta are \textit{paired}. In that case the
scaling goes as $s^{0}$, because now the delta function depends only on the
sum of momenta orthogonal to the Fermi surface and due to 
\begin{equation}
\delta (s l) \to s^{-1} \delta(l)
\end{equation}
the 4-fermi interaction indeed scales as $s^0$ and is marginal. (This
encapsulates the usual Cooper pairing phenomenon.) A one loop calculation of
the beta function does ensure the asymptotic freedom provided the
interaction is perturbatively attractive, as is the case for phonon-electron
interaction, which leads to the strong coupling regime (with bound states
forming) in the infrared. The resulting wave-function describing the
superconducting state is of a BCS kind 
\begin{equation}
\Psi_{BCS} \sim \prod_k ( \alpha + \beta a^*_k {a^*}_{-k}) \Psi_{FL}
\end{equation}
where $\alpha$ and $\beta$ are the usual variational parameters and $a^{*}$
denote the creation operators of the electron quasiparticle.

\subsection{ The Fermi liquid coupled to effective gravity}

In what follows we outline an intuitive argument for the crucial relevance
of an effective gravitational description of the normal state of the high $%
T_{C}$ materials. How could effective gravity arise from the dynamics of the
Fermi surface? First, we note that experiments indicate a highly irregular
Fermi surface in the normal state, resulting from the microscopic
physics \cite{htcreview}. This irregularity in turn could lead to an effective
gravitational description. As a concrete model, let us view the Fermi
surface as an incompressible fluid in momentum space. Then in analogy with
the discussion of induced gravity in fluid dynamics (\cite{unruh}, \cite{abh}%
) we can envision generating an effective metric, which in the case of real
irrotational fluids is precisely the acoustic metric \cite{unruh}. This
effective metric is generated from the fluctuations of the fluid density $%
\rho $ and the velocity potential $\phi $ (where the velocity $\vec{v}%
=\nabla \phi $). The underlying space-time action of the moving fluid is 
\begin{equation}
S=\int dx^{4}[\rho \dot{\phi}+\frac{1}{2}\rho (\nabla \phi )^{2}+U(\rho )]
\end{equation}%
with $U(\rho )$ denoting the effective potential. Upon variation of this
action one obtains the equations of motion for $\rho $ and $\phi $ (the
Euler continuity equation and the Bernoulli energy balance equation). When
these equations of motion are perturbed around the equilibrium values $\rho
_{0}$ and $\phi _{0}$ one is led to the equations for the fluctuations of
the velocity potential $\varphi $. In particular, the equation for the
fluctuations of the velocity potential can be written in a geometric form 
\cite{unruh} 
\begin{equation}
\frac{1}{\sqrt{-g}}\partial _{a}(\sqrt{-g}g^{ab}\partial _{b}\varphi )=0.
\end{equation}%
The effective space-time metric has the canonical ADM form of \cite{unruh}
and \cite{abh} (apart from a conformal factor) and has the Lorentzian
signature 
\begin{equation}
ds^{2}=\frac{\rho _{0}}{c}[c^{2}dt^{2}-\delta
_{ij}(dx^{i}-v^{i}dt)(dx^{j}-v^{j}dt)]
\end{equation}%
where $c$ is the sound velocity\footnote{%
or a plasmon velocity for a system of charged fermions} and $v^{i}$ are the
components of the fluid's velocity vector.

Note that the collective dynamics of the Fermi surface can be considered as
a \textquotedblleft bosonization\textquotedblright\ of the Fermi liquid \cite%
{boson} and in that approach the quasiparticle excitations can be
represented by a collective mode of the \textquotedblleft
bosonized\textquotedblright\ Fermi liquid. That approach runs into trouble
with the essential difference between fermions and bosons in spatial
dimensions one (where the bosonization can be used because of the very
special kinematics) and spatial dimensions above one, where such efforts are
largely prohibitive\footnote{%
For a sample of the literature on higher dimensional bosonizations please
consult \cite{boson}.}. In contrast, in our proposal the non-Fermi liquid
behavior originates from the gravitational dressing, itself caused by the
non-trivial geometry and topology of the Fermi surface, an experimental
fact. Thus we argue that the new quasiparticles are the usual fermions, but
propagating in a non-trivial gravitational background and hence dressed by
the gravitational fluctuations. In other words, the collective motion of the
Fermi surface is of an effective gravitational kind (i.e. not spin 0 but
spin 2) and the usual fermionic quasiparticles are now coupled to this
collective spin 2 bosonic mode. Thus we propose that the effective theory of
the strange normal state of high $T_{C}$ superconductors is a
gravitationally dressed Fermi liquid, i.e. the usual Fermi liquid albeit
coupled to gravity 
\begin{equation}
S_{NFL}=\int \sqrt{-g}L_{FL}(\psi ,\nabla \psi ).
\end{equation}

Applying this approach to the Fermi surface in momentum space, by relying on
the induced effective diffeomorphism invariance, we are led to conclude that
the effective action for the fermionic quasiparticle around the Fermi
surface should be a gravitationally dressed action, in which the naive
scaling dimensions discussed above can be changed by adding gravitational
dressing. In general this would mean that instead of $\psi \to s^{-1/2} \psi$
as discussed above, we should have 
\begin{equation}
\psi \to s^{-1/2 +\alpha} \psi
\end{equation}
where $\alpha$ denotes gravitational dressing\footnote{%
Such dressing can be explicitly computed in simple cases such as the
coupling of 2d gravity to 2d matter as discussed in \cite{kpz}.}. Thus the
fermionic two point function can be changed to scale as $s^{-1 +2\alpha}$,
indicative of a non-Fermi liquid behavior.

Naturally, the anomalous dimensions of Fermi propagators have been
considered previously \cite{htcreview, marginal, pw} \footnote{%
in the marginal Fermi liquid theory, through couplings with an effective,
induced gauge field, in the context of quantum critical fixed points, and
many other proposals \cite{htcreview}}, and our proposal is not unique in
placing\ emphasis on the anomalous fermionic propagators. Also, the
phenomenology implied by anomalous fermionic propagators, including
non-Fermi liquid behavior such as a resisitivity linear in temperature $T$ 
\cite{marginal} is inherent to our proposal as well. But the proposal that
the anomalous nature of the fermionic two-point function follows from the
concept of a normal state based on a gravitationally dressed Fermi liquid
Lagrangian describing the anomalous normal state is, to our knowledge, new.

In this scenario, the normal state is described by a gravitationally dressed
Fermi-liquid theory. Similarly, the BCS wave-function would be
gravitationally dressed. \textit{Note that our discussion is couched in
terms of an effective field theory and therefore does not carry information
about the underlying microscopic mechanism.} In particular, the specific
nature of the pairing mechanism and its relationship with the effective
gravitational description is outside the scope of the present approach. The
approach generally does conclude that the usual effective Landau-Ginsburg
theory describing the physics of the condensate is gravitationally dressed
as follows 
\begin{equation}
S_{LG_{g}}=\int d^{3}x\sqrt{-g}[g^{\mu \nu }\nabla _{\mu }H^{\ast }\nabla
_{\nu }H-V(|H|^{2})]
\end{equation}%
where the complex order parameter is denoted by $H$ and $V(|H|^{2})$ is an
effective potential (for example, of a quartic type). In what follows, most
important\ is that such a gravitationally dressed Landau-Ginsburg theory
leads to a simple geometric mechanism for the high $T_{C}$ of high
temperature superconductors.

\section{Effective gravity and energy scale hierarchy}

To implement a condensed matter analog of the gravitationally induced energy
scale hierarchy, we first present a brief summary of the Randall-Sundrum
proposal \cite{rs} which precisely accomplishes this goal. The exponential
hierarchy is induced by a warped extra dimensional geometry. The actual
example envisions two three-branes (i.e. two $3+1$ dimensional worlds),
separated by a distance of $r_{c}$ in one extra (fifth) spatial dimension.
This bulk $4+1$ dimensional space represents a portion of the $4+1$
dimensional AdS space, whose cosmological constant is determined by another
parameter of the model, $k$. One of the branes carries the degrees of
freedom of the Standard Model. The other carries the Planck scale degrees of
freedom.

This set-up can be carried over from the $3+1$ to the $2+1$ dimensional
context. The $2+1$ dimensional context carries relevance, because the high $%
T_{C}$ materials are layered anisotropic compounds where the Cu or Fe planes
can be identified with $2+1$ dimensional branes. The extra anisotropic
dimension is to be identified with the actual anisotropy of these materials.
The crucial assumption in this set-up is that effective gravity is induced,
perhaps by the internal dynamics of geometrically very complicated Fermi
surfaces. The warped geometry of the extra dimension could originate from
the inter-layer coupling. We address the question of how the effective
gravitational description could be induced from the dynamics of the Fermi
surface in Section 4.

Returning to the Randall-Sundrum scenario in the condensed matter situation,
the $3+1$ bulk metric reads 
\begin{equation}
ds^{2}=r_{c}^{2}du^{2}+e^{-2kr_{c}u}\eta _{\mu \nu }dx^{\mu }dx^{\nu }.
\end{equation}%
Here $u$ denotes the extra anisotropic spatial direction of size $l_{c}$,
and $\mu ,\nu =0,1,2$ are the planar space-time indices. Also, $\eta _{\mu
\nu }$ is the flat planar space-time (Minkowski) metric. The gravitational
fluctuations $h_{\mu \nu }$ around the flat Minkowski metric define the
background metric $g_{\mu \nu }^{0}=\eta _{\mu \nu }+h_{\mu \nu }$ \cite{rs}%
. The low-energy effective action for a Higgs field (order parameter), i.e.
the Landau-Ginsburg action, coupled to effective gravity is 
\begin{equation}
S_{LG_{g}}=\int d^{3}x\sqrt{-g}[g^{\mu \nu }\nabla _{\mu }H^{\ast }\nabla
_{\nu }H-\lambda (|H|^{2}-v_{0}^{2})^{2}]
\end{equation}%
where $g_{\mu \nu }=e^{-2kr_{c}l_{c}}g_{\mu \nu }^{0}$ \cite{rs} and $g$ is
the determinant of $g_{\mu \nu }$. After rescaling the Higgs field (the
Landau-Ginsburg order parameter) $H\rightarrow e^{kr_{c}l_{c}}H$ the
effective action becomes 
\begin{equation}
S_{LG_{g}}=\int d^{3}x\sqrt{-g^{0}}[g^{0\mu \nu }\nabla _{\mu }H^{\ast
}\nabla _{\nu }H-\lambda (|H|^{2}-e^{-2kr_{c}l_{c}}v_{0}^{2})^{2}].
\end{equation}%
Note that we can take $l_{c}=\pi $ as in \cite{rs}. Thus the physical mass
scale is given by a blue-shifted symmetry-breaking scale (where $v_{0}$ is
the minimum of the potential) 
\begin{equation}
v=v_{0}e^{-kr_{c}\pi }
\end{equation}%
or alternatively, the relevant mass scales are related as 
\begin{equation}
m=m_{0}e^{-kr_{c}\pi }.
\end{equation}%
This is the key result. Thus in order to obtain $m_{0}\sim 10^{19}GeV$ and $%
m\sim 1TeV$ one only needs to require $kr_{c}\sim 50$. Note that even though
this result has been derived in an effective field theory, there exists an
ultraviolet (UV) completion of this scheme from the point of view of string
theory \cite{joers}\footnote{%
In particular, the complex issue of stability of the Randall-Sundrum
scenario requires a UV complete description.}.

The question arises why the Randall-Sundrum warp factor should occur in the
interlayer coupling. We notice that the physics of interlayer tunneling
considered in the past \cite{pw} naturally involves exponential wave
functions and thus can be directly compared to the geometric exponential
factors that appear in the the Randall-Sundrum case. Thus the proposed
geometrization captures the relevant interlayer physics, at least on the
level of an effective, and not microscopic description. Here we emphasize
the weak nature of the inter-layer coupling as opposed to the strong intra-layer
coupling, the latter leading to non-Fermi liquid behavior and induced
gravitational dressing as argued in the previous section. Finally we remark
that in principle the geometric argument applies to both cuprates and
oxypnictides, and indeed stresses the very features they bear in common. 

\section{Why high $T_c$?}

Here we present a simple computation of $T_{C}$ based on the gravitationally
dressed Landau-Ginsburg effective description and the geometric argument
presented above. This constitutes the main result of this note.

Remembering the usual BCS dispersion relation given a gap $\Delta $ 
\begin{equation}
E=\sqrt{\Delta ^{2}+k^{2}}
\end{equation}%
in which the gap plays the role of the Higgs mass (given the well-known
analogy with the relativistic dispersion relation for a particle of mass $m$%
, i.e. $E=\sqrt{m^{2}+k^{2}}$), we see that there exists a distinct
possibility of similarly warping the gap, as in Eq. (5), and thus warping
the $T_{C}$.

Normally $T_{C}$ is related to the gap evaluated at low momenta \cite%
{feynman} 
\begin{equation}
T_{C}=const\Delta (0).
\end{equation}%
In the superconductors considered, one obtains a geometrically shifted
temperature so that the following relation results 
\begin{equation}
T_{HT_{c}}=T_{C}e^{kr_{c}\pi }.
\end{equation}%
Here $T_{C}$ denotes the low critical temperature of conventional
superconductors - typically a few K. $T_{HT_{c}}$ denotes the critical
temperature of high $T_{C}$ superconductors - equal to around $50-150$ K.

Therefore, a high $T_{C}$ naturally arises from the above exponential factor
that is responsible for a hierarchy of energy scales. An exponential factor
multiplying the usual 1-10 K critical temperature (determined by the value
of the BCS-like gap at zero momenta), produces a new critical temperature in
the 50-150 K range, or higher. A representative sample of, from top to
bottom, cuprate, oxypnictide (Fe-As) \cite{nature}, BCS type II, and BCS type I
superconducting critical temperatures is collected in the table below. For
example, transitioning from the lowest (Al) to the highest (Tl compound)
requires only an argument of 5 in the exponent.

\vskip .5cm

\begin{tabular}{|c|c|}
\hline
Compound & T$_{{C}}$ (K) \\ \hline
Tl$_{2}$Ba$_{2}$Ca$_{2}$Cu$_{3}$O$_{10}$ & 125 \\ \hline
Bi$_{2}$Sr$_{2}$Ca$_{2}$Cu$_{3}$O$_{10}$ & 110 \\ \hline
YBa$_{2}$Cu$_{3}$O$_{7}$ & 92 \\ \hline
La$_{2-x}$Sr$_{x}$CuO$_{4}$ & 38 \\ \hline
SmFeAsO$_{0.85}$F$_{0.15}$ & 42 \\ \hline
CeFeAsO$_{0.84}$F$_{0.16}$ & 41 \\ \hline
LaFeAsO$_{0.89}$F$_{0.11}$ & 26 \\ \hline
Nb$_{3}$Sn & 18 \\ \hline
NbTi & 10 \\ \hline
Pb & 7.2 \\ \hline
Al & 1.1 \\ \hline
\end{tabular}

\section{Conclusions}

In this note we have proposed a condensed matter analog of the
Randall-Sundrum scenario for a geometrically induced hierarchy of scales. We
have argued that in the context of the physics of high temperature
superconductivity, such a scenario can be responsible for the high $T_{C}$
of cuprate and oxypnictide superconductors. The conjecture spontaneously
leads to emphasis on the unusual features shared by cuprates and
oxypnictides. Needless to say we have only scratched the surface of the
manifold of issues relevant for high $T_{C}$ superconductivity. Of primary
importance remain a specific microscopic mechanism as well as a more
detailed phenomenology derived from the gravitationally dressed
Landau-Ginsburg approach. Apart from advancing understanding of the unusual
normal as well as the superconducting phase, this new approach can
illuminate other parts of the phase diagrams of high temperature
superconductors. For instance, of future interest are the relevance of
spin-spin interaction \cite{pm} for the effective gravitational description
proposed in this note, a study whether the $d$-wave nature of the order
parameter \cite{htcreview} naturally follows from a gravitationally dressed
Landau-Ginsburg description, an understanding of the pseudo-gap region of
the phase diagram, and finally connecting the recent discussion of Mottness 
\cite{mottness} to our context. These and other questions are left for more
detailed future work.

\vskip .5cm

\textbf{\Large Acknowledgements}

\vskip .5cm

We would like to thank R. G. Leigh, Philip Phillips, Joseph Polchinski and
Victoria Soghomonian for lending their ears to a preliminary version of this
proposal. Many thanks to Laurent Freidel, Vishnu Jejjala and Nemanja Kaloper
for comments on the draft of this note. {\small DM} is supported in part by
the U.S. Department of Energy under contract DE-FG05-92ER40677. {\small JJH}
is supported in part by the National Science Foundation under contract
DMR-0618235.

\end{document}